\documentclass[comment]{epl2}

\usepackage{amsmath, amssymb}
\usepackage{multirow}
\usepackage{booktabs}
\usepackage{hyperref}
\usepackage[inline]{enumitem}

\title{Comment on ``Measurements of Newton's gravitational constant and the length of day''}

\author{M. Pitkin}

\institute{
  SUPA, School of Physics \& Astronomy, University of Glasgow, Glasgow, G12 8QQ, UK
}


\begin{document}

\maketitle

\section{Introduction}

In \cite{2015EL....11010002A} the authors claim to observe a periodic signal in measurements of
Newton's gravitational constant, $G$. Specifically they find a 5.9 year period signal that
is strongly correlated with variations in the observed length of day
\cite{2013Natur.499..202H}. They do not suggest that $G$ actually varies on these
time-scales, but rather that there could be some systematic effect on the measurement process that is
correlated with the mechanism that leads to the variation in the length of day.
Here I present a reanalysis of the data used in \cite{2015EL....11010002A}
using Bayesian model selection to test the hypothesis that the data contains
a periodic signal compared to other potential models. In light of updated information on the times
of the various $G$ measurements given in \cite{2015arXiv150501774S} I also reanalyse this new dataset with
the same method. In both datasets I have found that a model for the variations in $G$ that only contains an
additional Gaussian noise term is hugely favoured, by factors of $\gtrsim e^{30}$, over models containing
a sinusoid term.\footnote{The code, data tables, figures and prior ranges for this analysis can be
found at \url{https://github.com/mattpitkin/periodicG}.}

\section{Analysis method}

Bayesian model selection provides a natural way to test multiple hypotheses by forming the odds ratio of
evidences for the different hypotheses. The odds ratio for two hypotheses $H_i$ and $H_j$ is given by
\begin{equation}
 \mathcal{O}_{ij} = \left(p(d|H_i,I)/p(d|H_i,I)\right)\times\left(p(H_i|I))/(p(H_j|I)\right)
\end{equation}
where $p(d|H_i,I)$ is the evidence for hypothesis $H_i$ given some data $d$,
$p(H_i|I)$ is the prior probability for $H_i$, and $I$ is information concerning any other assumptions.
When comparing hypotheses I assume that they are equally
probable a priori, so the prior ratio is unity. Therefore, I just calculate
the ratio of evidences for each hypothesis. If a given hypothesis is defined
by a set of parameters, $\vect{\theta}_i$, with their own priors,
$p(\vect{\theta}_i|H_i,I)$, then to calculate the evidence the parameters must be marginalised
over, e.g.\
\begin{equation}\label{eq:evidence}
 p(d|H_i,I) = \int^{\vect{\theta}_i} p(d|\vect{\theta}_i,H_i,I) p(\vect{\theta}_i|H_i,I) {\rm d}\vect{\theta}_i,
\end{equation}
where $p(d|\vect{\theta}_i,H_i,I)$ is the likelihood function of the data given a set of model parameters $\vect{\theta}_i$.

The general model that I use for my hypotheses is
\begin{equation}\label{eq:model}
 m(\mu_G, A, P, \phi_0, T_k) = A\sin{(\phi_0 + 2\pi (T_k-t_0)/P)} + \mu_G,
\end{equation}
where $\mu_G$ is an offset value, $A$ is the sinusoid amplitude, $\phi_0$ is an initial phase at an epoch $t_0$,
$P$ is the sinusoid period, and $T_k$ is the time.

In this analysis I have compared four different hypotheses, $H_i$, to explain the measurements of $G$:
\begin{enumerate*}[label=$H_{\arabic*}$\upshape)]
 \item the data is consistent with Gaussian errors, given by the experimental error bars $\sigma_{e,k}$, about an 
 unknown $\mu_G$;
 \item as for $H_1$, but also including an unknown common Gaussian noise term $\sigma_{\mathrm{sys}}$;
 \item as for $H_1$, but also including a sinusoid with unknown $A$, $\phi_0$ and $P$; and, 
 \item as for $H_3$, but also including an unknown $\sigma_{\mathrm{sys}}$.
\end{enumerate*}
These each correspond to a different set of parameters required in $\vect{\theta}$ and also the number of
parameters required in the integral of eq.~\ref{eq:evidence}.

For an initial examination of the claim in \cite{2015EL....11010002A} I have used their Figure~1 to read-off
the experimental times and then used Table~XVII of \cite{RevModPhys.84.1527} for values of $G$\footnote{For
the BIPM-13 measurements I used the combined servo and Cavendish value from \cite{PhysRevLett.113.039901}
and for the LENS-14 measurements I used the values from \cite{2014Natur.510..518R}.}.

In \cite{2015EL....11010002A} the experiment times are given no associated error.
However, many of the times used correspond to the received date of the respective paper rather than the
date of the actual experiment.  In analysing this data I specified uncertainties on the experiment times of
$\sigma_{t} = 0.25$\,years (with the exception of the JILA-10 and LENS-14 measurements for which I use uncertainties
of one week) before the given time. I have taken this time uncertainty into account by marginalising over
it for each data point.

\section{Results}

The odds ratios comparing hypotheses when using the $G$ dataset of \cite{2015EL....11010002A} are summarised in
table~\ref{tab:results}. It is clear that hypotheses including extra parameters over that for $H_1$
are hugely favoured by factors of $\gtrsim e^{100}$. The two hypotheses, $H_3$
and $H_4$, containing a sinusoidal signal are both approximately equally probable. However, $H_2$, just
containing the additional unknown noise term $\sigma_{\mathrm{sys}}$ and the unknown offset $\mu_G$, is
hugely favoured by factors $\sim e^{30}$
over $H_3$ and $H_4$. This shows that the simple model for which variations are just due to an unknown Gaussian
noise term is far more likely to be the cause of the variations than an additional sinusoidal variation.
This is due to Bayesian model selection naturally applying a penalty for including additional parameters that
do not significantly increase the evidence.

\begin{table}
\caption{Log odds ratios for the four hypotheses ($i$ represents rows and $j$ represents columns)
when using the data used in \cite{2015EL....11010002A}, with those when using the data from
\cite{2015arXiv150501774S} in parentheses.}
\label{tab:results}
\begin{center}
 \begin{tabular}{c|ccc}
  $\ln{\mathcal{O}_{ij}}$ & $H_2$ & $H_3$ & $H_4$ \\
  \specialrule{0.25pt}{0.75pt}{0.75pt}
  $H_1$ & $-133$ ($-140$) & $-102.2$ ($-66$) & $-103$ ($-110$) \\ 
  $H_2$ &  & $30$ ($74$) & $30$ ($30$) \\
  $H_3$ &  &  & $-0.3$ ($-45$)
 \end{tabular}
\end{center}
\end{table}

I have also looked at the posterior probability distributions for the period and for $H_3$ I see a
clear lone spike in probability around the claimed period of 5.9\,years. A similar spike
shows up for $H_4$, but is much less pronounced.
I have assessed the significance of this period probability peak for $H_3$ by rerunning that analysis 20 times,
but each time randomly shuffling the $G$ values to remove any real periodicity in the data.  Out of these 20 runs
there is one time when the hypothesis using the shuffled data
is more favoured than when using the un-shuffled data and another couple that are within a factor of two.
The posteriors for these cases also show very similar spikes in the period to that from the unshuffled data.

Since the acceptance of \cite{2015EL....11010002A} Schlamminger {\it et al.} \cite{2015arXiv150501774S}
examined the claim, in particular noting that the experimental times in the original work
are not accurate. They examined the literature to compile a more complete list of
experiments with information on the actual dates on which the experiments were performed.
I have reanalysed this new dataset for each of the four hypotheses. When marginalising over the
time error I have now set the error to be symmetric around the mean experiment times.
For all other parameters I have used the same prior ranges as in the initial analysis.
The odds ratios for each of these cases are also given in table~\ref{tab:results} from which it can
be seen that $H_2$ is still favoured over all other hypotheses by a huge amount. However, $H_3$
is now hugely disfavoured over $H_4$, i.e.\ just including a sinusoid, but adding no additional noise
term does far worse at fitting the data than also including the noise term.

\section{Conclusions}

I have reanalysed the data consisting of measurements of $G$ from
\cite{2015EL....11010002A} and \cite{2015arXiv150501774S} to asses the claim of a periodic
component with a period of 5.9\,years.

Using Bayesian model selection, and four different hypotheses to describe the variations in the data, and including
uncertainties on the experimental times, I have found that the best model is one in which
there is an additional unknown Gaussian noise term on top of the observed experimental errors. This is 
favoured over a model also containing a sinusoidal term by factors of $\gtrsim e^{30}$. I also find that periodic
signals can easily be found in random permutations of the data suggesting that the observed periodicity seen in
\cite{2015EL....11010002A} is just a random artifact of the data.

Following the publication of \cite{2015EL....11010002A} the authors have taken into account the work of \cite{2015arXiv150501774S} (see \cite{AndersonRevised}). They fit an additional
sinusoid to the updated data and note that the significance of the correlation with the length of day decreases. 
I expect that calculating the evidence for a model including two sinusoids would not cause such a model to be
favoured over the simpler model containing just the
extra Gaussian noise term, as the increase in parameter space will be penalised if the fit does not 
significantly improve.

I note that if there were good a priori reasons to expect a periodic component with a specific period in the data
(i.e.\ if there were a good reason why the mechanism leading to changes in the length of day could couple into 
measurements of
$G$), then the evidence for models containing such a periodic signal might dramatically increase. However, without such
prior knowledge using such a constraint would strongly bias us.

\acknowledgements

I would like to thank Prof.\ J.\ Faller for useful discussions, Prof.\ C.\ Speake for putting me in contact with
Dr.\ S.\ Schlamminger, and Dr.\ Schlamminger for providing me with a data file of their compiled $G$ measurements.
I am funded by the STFC under grant ST/L000946/1.
~\\
~\\
{\it Additional remark}: In responding to my Comment \cite{response} the authors of the original article put forward two related pieces of 
evidence to suggest that models for the time variations of $G$ measurements containing a constant 
offset and one or two periodic components are favoured over a model with no periodic component. 
They show that once the best fits for the models containing one or two periodic components are 
removed from the data the residuals have smaller variance, and have a distribution that is closer 
to a Gaussian distribution, than for the model containing no periodic component. These findings are 
not at all surprising.  If one adds more parameters to the model, then one can almost always find a 
better fit with smaller residuals -- in cases where the additional model complexity adds no extra 
information this is commonly known as over-fitting and is related to the concept of Occam's razor. 
A Bayesian analysis, such as I performed, naturally allows one to penalise extra complexity when it 
is unnecessary through the incorporation of an Occam factor. This Occam factor penalty comes 
about through use of the prior volume of the parameter space for each model. Each parameter has a 
range over which it is {\it a priori} thought to exist, and in my analysis the prior volume is 
just the product of these ranges. So, a model with more parameters will often have a larger prior 
volume. This naturally incorporates a penalty for over-fitting in that larger prior volumes will 
down-weight the evidence for a model, so if the likelihood does not compensate enough for this 
down-weighting then the evidence will be reduced.

In my analysis this penalty far outweighs the slightly better fit that can be achieved with a more 
complex model containing a sinusoid. It should also be noted that my most favoured model does not 
just contain a constant offset, but also contains an additional unknown noise term. This may 
suggest that the quoted errors on the $G$ measurements are underestimates of the true noise.

A further addition that I was able to include in my analysis, but which the authors of the original 
article do not appear to have addressed is that there are also error bars on the
experimental times. This may also weaken their fit.

Further $G$ measurement data may change my conclusions, but with the current data I stand by my 
result that the data is best explained by just a constant offset and an additional unknown 
noise term, and no periodic component is required.

\bibliographystyle{eplbib}
\bibliography{comment_revised}

\end{document}